\documentclass[a4paper,11pt]{article}

\usepackage{pos}

\usepackage[
  style=numeric-comp,
  citestyle=numeric-comp,
  backend=biber,
  sorting=none,
  doi=false,
  isbn=false,
  url=false,
  eprint=true,
  giveninits=true
]{biblatex}

\DeclareSourcemap{
  \maps[datatype=bibtex]{
    \map{
      \step[fieldsource=arxivid, fieldtarget=eprint]
      \step[fieldset=eprinttype, fieldvalue=arxiv]
    }
  }
}

\DeclareNameAlias{default}{given-family}

\DeclareFieldFormat[article]{title}{\mkbibemph{#1}}
\DeclareFieldFormat[inproceedings]{title}{\mkbibemph{#1}}

\DeclareFieldFormat{volume}{#1}
\DeclareFieldFormat{number}{#1}
\DeclareFieldFormat{pages}{#1}

\DeclareFieldFormat{eprint:arxiv}{
  \mkbibbrackets{\href{https://arxiv.org/abs/#1}{#1}}
}

\DeclareFieldFormat[article]{journaltitle}{
  \iffieldundef{doi}
    {\mkbibemph{#1}}
    {\mkbibemph{\href{https://doi.org/\thefield{doi}}{#1}}}
}

\renewbibmacro{in:}{%
  \ifentrytype{article}{}{\printtext{\bibstring{in}\intitlepunct}}}

\renewcommand{\logo}{\relax} 

\newcommand{\diracgamma}{\gamma^j}

\newcommand{\inprod}[2]{\langle{#1},{#2}\rangle}

\newcommand{\R}{\mathbb{R}}
\newcommand{\C}{\mathbb{C}}
\newcommand{\HH}{\mathbb{H}}
\newcommand{\Z}{\mathbb{Z}}

\newcommand{\hilbert}{\mathcal{H}}
\newcommand{\A}{\mathcal{A}}

\newcommand{\trigamma}[3]{\gamma^{#1}\gamma^{#2}\gamma^{#3}}

\newcommand{\comm}[2]{\left[{#1},{#2}\right]}
\newcommand{\acomm}[2]{\{#1,#2\}}

\DeclareMathOperator\Tr{Tr}

\newcommand{\U}{{\mathrm U}}
\newcommand{\PU}{{\mathrm {PU}}}
\newcommand{\OO}{{\mathrm O}}      

\newcommand{\PO}{{\mathrm{PO}}}
\newcommand{\Sp}{{\mathrm{Sp}}}
\newcommand{\PSp}{{\mathrm{PSp}}}

\title{Fuzzy Geometries with an Internal Space}

\author{John W. Barrett}
\author*{Joseph Burridge}

\affiliation{School of Mathematical Sciences, University of Nottingham,\\
University Park,
Nottingham NG7 2RD, United Kingdom}

\emailAdd{John.Barrett@nottingham.ac.uk}
\emailAdd{Joseph.Burridge@nottingham.ac.uk}

\abstract{The product of a non-commutative matrix spectral triple with a simple two-dimensional internal space is considered. This is interpreted as a non-commutative spacetime that contains one charged Dirac fermion and its antiparticle. The inner fluctuations of a vacuum Dirac operator are calculated, using the standard technique of Connes' one-forms. This results in the non-commutative analogue of a gauge field, as expected, and also fluctuations of the spacetime geometry. In addition, the fluctuations include a derivative operator that depends on the particle charge. The integral over the fermions in the model is calculated, leading to some novel induced bosonic terms.
}

\FullConference{Proceedings of the Corfu Summer Institute 2025 "School and Workshops on Elementary Particle Physics and Gravity" (CORFU2025)\\
27 April - 28 September, 2025\\
Corfu, Greece\\}

\begin{document}
\maketitle

\section{Introduction}
Noncommutative geometry is the generalisation of geometry to spaces that can be described by a commutative or non-commutative algebra. A spectral triple is a particular type of noncommutative geometry. It encodes the mathematical properties of a Dirac operator, providing a framework for describing the geometry of both commutative spaces (Riemannian manifolds) and non-commutative spaces. 

The spectral triple approach has found success in relation to the Standard Model. In this approach, the Standard Model fields arise from an \emph{almost-commutative geometry}. This is a product of a commutative spectral triple, corresponding to spacetime, with a spectral triple corresponding to an internal space, that can be commutative or non-commutative. The internal space has a finite-dimensional Hilbert space with one basis vector for each type of fermion and its antiparticle. The geometry of the internal space governs the fermion masses and the transformations of the fermions under the gauge group. This construction provides a geometric explanation for the particle content of the SM \cite{Connes1995NoncommutativeReality,Connes1996GravityGeometry,Chamseddine2006GravityMixing}.

A natural extension of almost-commutative models is to replace the Riemannian manifold with a non-commutative manifold. This is motivated by attempts to discretise spacetime, model quantum gravitational effects, and examine how classical geometry may emerge from finite or fuzzy structures. These non-commutative manifolds come in many flavours; for example, Moyal plane-type spacetimes \cite{Moyal1949QuantumTheory,Gayral:2003dm}, the noncommutative torus \cite{Connes1997NoncommutativeTori}, matrix geometries \cite{Barrett2015MatrixTriples}, including particular examples such as the fuzzy sphere or the fuzzy torus \cite{Madore1992TheSphere,Barrett2025FiniteTorus}, or truncations of commutative spaces \cite{Glaser2021ReconstructingTriples, Connes2021SpectralSystems}.

Product geometries based on non-commutative manifolds have been explored in the case of matrix geometries. As the internal spaces of these products carry a natural gauge group, they provide a framework for formulating gauge theory on fuzzy manifolds via spectral triples. Examples in which both the manifold and the internal space are matrix geometries were constructed in
\cite{Perez-Sanchez2021OnModel}. The inner fluctuations of this model were analysed, and, via the spectral action, were shown to produce a non-commutative analogue of a Yang-Mills-Higgs model. 

This paper explores a different sort of internal space, one containing just two fermion fields: a $\U(1)$-charged fermion and its antiparticle, which has the opposite charge. An important feature is that the spectral triples are based on real algebras (algebras over $\R$), as is necessary for realistic particle physics. This means that the matrix geometry of the `spacetime' manifold is also described with a real algebra. This allows a certain refinement of the usual (complex algebra) description of a matrix geometry (\S\ref{sec:matrixgeometry}).

After taking the product with a matrix geometry, the model describes a non-commutative version of spacetime containing a Dirac fermion field and its antiparticle (\S\ref{sec:internal}). Note that the chiral version of this model (which is not considered here) would contain a right-handed electron and its antiparticle.

The space of inner fluctuations is explored for the model (\S\ref{sec:fluct}). It is found that the fluctuations include both universal fluctuations that couple to the particle and antiparticle identically, and charged fluctuations that couple according to the $\U(1)$ charge. The universal fluctuations simply perturb the Dirac operator of the spacetime. The charged fluctuations lead to an analogue of a $\U(1)$ gauge field, and another new field that is a derivative operator that couples to the charge of the fermion field. This latter type of fluctuation is a new phenomenon that appears to be linked to the non-locality of gauge transformations in a non-commutative spacetime.

The associated fermionic field theory is then examined by computing a fermionic integral (\S\ref{sec:ferm}). This results in a function of the bosonic fields that is interpreted as an induced action. This provides a framework for analysing fermionic loop effects in spectral triple models. For the fluctuations found, the resulting determinant can be described in terms of a generalisation of the field strength tensor.

\section{Matrix Geometries}\label{sec:matrixgeometry}
\subsection{Spectral triple}\label{sec:spectr}
A finite spectral triple $(\mathcal{A},\mathcal{H},D)$ consists of $\mathcal{A}$, a unital $*$-algebra, a faithful $*$-representation $l$ of $\mathcal A$ on a finite dimensional Hilbert space $\mathcal{H}$, and a Hermitian operator $D:\mathcal{H}\rightarrow \mathcal{H}$. To form a finite real spectral triple, a $\mathbb{Z}_2$ grading $\Gamma$ (i.e $\Gamma^*=\Gamma$, $\Gamma^2=I$) called the chirality operator is introduced on $\mathcal{H}$, as well as a real structure $J$. These are analogous to the chirality $\gamma$ and real structure $C$ on the Hilbert space of spinors at a point on a spin manifold. The data for a finite real spectral triple satisfies the axioms given in \cite{Barrett2015MatrixTriples}. There are eight different types of real structure, determined by a parameter $s$ known as the KO-dimension of the spectral triple.

In a real spectral triple, there is a right action of $\mathcal A$ on $\mathcal H$ determined by
\begin{equation*} r(a)=J\,l(a^*)\,J^{-1}.
\end{equation*}
This makes $\mathcal H$ a bimodule over $\mathcal A$. It is convenient to denote $l(a)\psi-r(a)\psi=[a,\psi]$ and $l(a)\psi+r(a)\psi=\{a,\psi\}$ for any $\psi\in\mathcal H$. In places, this notation will be extended to elements of $\mathcal A\otimes \C$ by extending $l$ and $r$ to be complex-linear.
The commutator is a non-commutative analogue of a derivative, and the anti-commutator is the non-commutative analogue of the multiplication of a spinor field by a function \cite{BarrettIntro}. 

A \emph{matrix geometry} is a particular type of finite spectral triple, where the algebra is assumed to be a simple matrix algebra \cite{Barrett2015MatrixTriples}. The Hilbert space is
$\hilbert =V \otimes M_n(\C)$, where
$V$ is a module for a real Clifford algebra of type $(p,q)$  (having $p$ generators squaring to $1$ and $q$ squaring to $-1$). This Clifford module has a chirality operator $\gamma$ and a real structure $C$. 

The general construction for a matrix geometry is as follows 
\begin{itemize}
    \item $s \equiv q-p \mod{8}$, the KO-dimension
    \item $\mathcal{A} = M_n(\mathbb{R}),M_n(\mathbb{C}),\text{ or } M_{n/2}(\mathbb{H})$, all considered as real algebras
    \item $\mathcal{H}  = V \otimes M_n(\mathbb{C})$, a complex vector space
    \item An inner product on $\mathcal H$: $\langle v\otimes m, w \otimes m'\rangle = (v,w)_V\Tr(m^*m')$
    \item A representation of $\mathcal{A}$ on the Hilbert space: $l(a)(v\otimes m) = v \otimes am$
    \item A chirality operator: $\Gamma (v\otimes m) = (\gamma v) \otimes m$
    \item A real structure: $J(v\otimes m) = Cv \otimes m^*$.
\end{itemize}
Note that the above data without a Dirac operator defines a \emph{fermion space}. The vector space of all possible Dirac operators, $\mathcal{G} = \{D\}$, is the space of geometries for the given fermion space. In general, it is determined by the real linear conditions
\begin{itemize}
\item $D=D^*$
\item $D=\epsilon' JDJ^{-1}$,
with $\epsilon'=\pm1$ determined by $s$
\item $[[D,l(a)],r(b)]=0$ for all $a,b\in\mathcal A$ (the first-order condition)
\item For even $s$, $D\Gamma+\Gamma D=0$.
\end{itemize}
For a matrix geometry, Dirac operators can be written in an explicit form determined by products of gamma matrices and operators on the matrix Hilbert space that are commutators or anti-commutators \cite{Barrett2015MatrixTriples}. In this paper, the type $(0,4)$ matrix geometry is used as the main example. This is because the formulas are direct analogues of the Dirac operator for a Riemannian 4-manifold. The formulas for type $(0,n)$ are similar, but simplify for $n<4$. For $n>4$ there may be products of 5 or more gamma matrices, which are not of interest for physical applications.

A $(0,4)$ matrix geometry has the Dirac operator
\begin{equation}
    D = \sum_{j} \gamma^{j}
     \comm{L_{j}}{\cdot\,} + \sum_{
     j<k<l} \gamma^j\gamma^k\gamma^l  \acomm{H_{jkl}}{\cdot\,} \label{eq:Dirac}
\end{equation}
for some $L_{j}, H_{jkl}\in M_n(\C)$. The gamma matrices $\gamma^j$ act in $V$, but the same notation is used for the action in $V\otimes M_n(\C)$ (strictly, this is $\gamma^j\otimes 1$).
Due to the fact that the gamma matrices are anti-Hermitian, the matrices $L_j$ are anti-Hermitian and the $H_{jkl}$ are Hermitian.

In the analogy with commutative geometry, the first term of the Dirac operator has a commutator and is thus an analogue of a derivative operator. The second term, with the anti-commutator, is analogous to a spin-connection term. 

Using a real algebra, $\mathcal A$, allows a refinement of this formula by requiring that 
$$L_{j}, H_{jkl}\in\A,$$ 
and this is done in this work. Note that in the case $\A=M_n(\C)$, this makes no difference, and so the Dirac operators are exactly those operators that satisfy the axioms \cite{Barrett2015MatrixTriples}. But in the cases $\mathcal A=M_n(\R)$ or $M_{n/2}(\HH)$ this requirement restricts to a smaller vector space of Dirac operators. This restriction is the analogue of the requirement that the derivatives and functions in a commutative Dirac operator have real coefficients.

\subsection{Inner Fluctuations}
Gauge theory is naturally defined on spectral triples via their inner automorphism group, generated by unitary elements of the algebra. The gauge group, $G(\mathcal{A},\mathcal{H};J)$, of a spectral triple is 
\begin{equation*}
    G(\mathcal{A},\mathcal{H};J) \colon= \left\{ U=uJuJ^{-1} : u\in \U(\mathcal{A}) \right\}.
\end{equation*}
The gauge group for $(p,q)$ matrix geometries depends upon the choice of matrix algebra. As an example, consider $\A = M_n(\R)$. The unitary elements of the algebra are
\begin{equation*}
    \U(\mathcal{A}) = \OO(n)
\end{equation*}
The subalgebra of elements that commute under the real structure is given by $\mathcal{A}_J = \mathbb{R}I_n$, with the unitary elements of this subalgebra being $\{\pm 1\}$. Therefore, the gauge group is the quotient
\begin{equation*}
     G(\mathcal{A},\mathcal{H};J) = \OO(n)/\{\pm 1\} = \PO(n) 
\end{equation*}
In a similar fashion, the gauge groups for $M_n(\C)$ and $M_{n/2}(\mathbb{H})$ are given by
$\U(n)/\U(1) = \PU(n)$ and $\Sp(n/2)/\{\pm 1\} = \PSp(n/2)$ (here $\Sp(n)$ is the compact symplectic group or quaternionic unitary group). 

The left inner fluctuations are defined using \emph{Connes one-forms}. These are Hermitian operators of the form
\begin{equation*}
        \omega = \sum_m l(a_m)\comm{D}{l(b_m)}.
\end{equation*}
The sum is taken over a finite set of pairs of elements $(a_m,b_m)\in \mathcal A\times\mathcal A$.
    
The fluctuation of the Dirac operator is the addition of a term
\begin{equation*}
\Omega= \omega+\epsilon'J\omega J^{-1}
\end{equation*}
with $\epsilon'=\pm1$ determined by the KO-dimension. As an example, the fluctuation determined by a unitary conjugation is
\begin{equation*}
\Omega=  UDU^{-1}-D=u[D,u^{-1}]+\epsilon'\, Ju[D,u^{-1}]J^{-1},
\end{equation*}
i.e., an inner fluctuation with $a=u$, $b=u^{-1}$, and $u\in\U(\A)$. These fluctuations generate an isospectral Dirac operator and are a non-commutative generalisation of a gauge transformation. In fact, in the almost-commutative case fluctuations of this type coincide with gauge transformations.

For an almost-commutative manifold, Connes one-forms correspond to de Rham one-forms, and with a suitable internal space, reproduce gauge fields on the manifold. The above definition of a Connes one-form is therefore used as the starting point for investigating these features on matrix geometries.

\subsection{Fluctuations of \texorpdfstring{$\varepsilon'=1$}{ε'=1} Matrix Geometries}\label{sec:fluctuations}
In the case of matrix geometries with $\varepsilon'=1$, of which the (0,4) geometry is an example, the Dirac operator takes a standard form
\begin{equation*}
    D = \alpha^\mu  \comm{L_\mu}{\cdot} + \rho^\nu \acomm{H_\nu}{\cdot},
\end{equation*}
where $\alpha^\mu$ are general anti-Hermitian products of gamma matrices, and $\rho^\nu$ are general Hermitian products. There is an implicit summation over the indices $\mu$ and $\nu$.

In this case, the left fluctuation is given by 
\begin{equation}
    \omega =\sum_m \alpha^\mu \, l(a_m\comm{L_\mu}{b_m}) + \rho^\nu \, l(a_m\comm{H_\nu}{b_m})\label{leftfluc}
\end{equation}

Recalling the Hermiticity constraint on $\omega$, the factor $\sum_m  l(a_m\comm{L_\mu}{b_m})$ is anti-Hermitian whilst $\sum_m  l(a_m\comm{H_\nu}{b_m})$ is Hermitian. 

Using the condition $J\diracgamma J^{-1} = \diracgamma$, the right fluctuation is
\begin{equation*}
    J\omega J^{-1} = \sum_m -\alpha^\mu \, r(a_m\comm{L_\mu}{b_m}) + \rho^\nu \, r(a_m\comm{H_\nu}{b_m})
\end{equation*}
and hence the total fluctuation is
\begin{equation*}
    \Omega = \sum_m \alpha^\mu \, \comm{a_m\comm{L_\mu}{b_m}}{\cdot} + \rho^\nu \, \acomm{a_m\comm{H_\nu}{b_m}}{\cdot}
\end{equation*}

The fluctuated Dirac operator is
\begin{align*}
    D'\equiv D +\Omega &= \sum_m \alpha^\mu \, \comm{L_\mu + a_m\comm{L_\mu}{b_m}}{\cdot} + \rho^\nu \, \acomm{H_\nu + a_m\comm{H_\nu}{b_m}}{\cdot} \\
    &= \alpha^\mu \, \comm{L_\mu'}{\cdot} + \rho^\nu \, \acomm{H_\nu'}{\cdot} 
\end{align*}
which is a simple redefinition of the original Dirac operator with new matrices $L'$ and $H'$. Therefore, these one-forms have the effect of globally shifting the geometry encoded by the original Dirac operator. The resulting orbits of Dirac operators under fluctuations appear to be complicated in general, but it can easily be observed that if any one of the $L_\mu$ or $H_\nu$ is zero, then it remains zero after the fluctuation. 

In the case of a unitary fluctuation, $\omega = u\comm{D}{u^*}$ where $u\in \U(\A)$, the modified Dirac operator is given by
\begin{align*}
    D + \Omega &= \alpha^\mu \, \comm{L_\mu + u\comm{L_\mu}{u^{-1}}}{\cdot} + \rho^\nu \, \acomm{H_\nu + u\comm{H_\nu}{u^{-1}}}{\cdot} \\
    &=\alpha^\mu \, \comm{uL_\mu u^{-1}}{\cdot} + \rho^\nu \, \acomm{uH_\nu u^{-1}}{\cdot} \\
    &= UDU^{-1}.
\end{align*}
This subset of fluctuations can be viewed as \emph{geometric gauge transformations}. In the analogy with commutative geometry, both the derivative operators and the spin connection coefficients change, although the geometry is isomorphic. Geometrically, these transformations correspond to area-preserving diffeomorphisms in simple cases \cite{Ishiki2019DiffeomorphismsSphere}. This differs from the commutative case, where there are no perturbations of the manifold.

\section{U(1) Internal Space}\label{sec:internal}
The product of spectral triples \cite{Vanhecke1999OnTriples, Dabrowski2010ProductTriples} generalises the Cartesian product of manifolds in the commutative context. In the case of an almost-commutative spectral triple, the fluctuated Dirac operator contains Connes one-forms that provide the gauge and Higgs field content for physics on a Riemannian manifold. In this paper, the commutative spectral triple is replaced with a matrix geometry. Thus, the model comprises a (0,4) matrix geometry tensored with an internal space, which is also a finite real spectral triple. 

The internal space in this paper is the $s=6$ spectral triple constructed as follows
\begin{equation*}
    \mathcal{A}_F = \C, \;  \hilbert_F = \C^2, \; l_F(\kappa) = \begin{pmatrix}
        \kappa & 0 \\ 0 & \overline{\kappa}
    \end{pmatrix}, \; D_F = 0
\end{equation*}
Here $\C$ is considered as a real algebra. It has real structure and chirality operator,
\begin{equation*}
    J_F\begin{pmatrix}e\\p \end{pmatrix}=\begin{pmatrix}\overline p\\\overline e \end{pmatrix} , \;  \Gamma_F = \begin{pmatrix}
        1 & 0 \\ 0 & -1
    \end{pmatrix},
\end{equation*}
and the Dirac operator is zero due to the first-order condition.
Physically, this internal spectral triple corresponds to a charged particle and its antiparticle, for which there is no Majorana mass term. It was shown to have gauge group U(1) in \cite{Bhowmick2011QuantumGeometry}, hence this internal triple is referred to as a \emph{U(1) internal space}. 
The internal space is simpler than the two-point internal space of \cite{Dungen2012ParticleSpacetimes} because in the real algebra formalism it is not necessary to double the internal space algebra to $\C\oplus\C$.

The matrix geometry from \S \ref{sec:spectr} is denoted here by the subscript $M$, so the spectral triple is $(\mathcal A_M,\mathcal H_M,\mathcal D_M)$. The algebra $\mathcal A_M$ is restricted to be either $M_n(\R)$ or $M_{n/2}(\HH)$ in the following. (The case $M_n(\C)$ is substantially different.)

The product space is the KO-dimension 2 triple
\begin{align*}
    \A &= \A_M\otimes\A_F=\A_M \otimes \C \\
    \hilbert &= \hilbert_M\otimes \hilbert_F= \C^4 \otimes M_n(\C) \otimes \C^2 \\
    J &= J_M\otimes J_F= C  \otimes (\cdot)^* \otimes J_F \\
    \Gamma &= \Gamma_M \otimes \Gamma_F = \gamma \otimes 1_n \otimes \Gamma_F ,
\end{align*} 
with vacuum Dirac operator $ D_0 = D_M \otimes 1_F $.
The representation of the algebra is determined by
    \begin{equation*}
        l(A\otimes\kappa) =  l_M(A) \otimes\begin{pmatrix}
            \kappa & 0 \\ 0 & \overline{\kappa}
        \end{pmatrix} = \begin{pmatrix}
            l_M(\kappa A) &0 \\ 0 & l_M(\overline{\kappa} A)
        \end{pmatrix}.
    \end{equation*}
An arbitrary element $a\in\mathcal A$ can be written $a=x+iy$, with $x,y\in \mathcal A_M$. Then the complex conjugate can be defined as $\overline a=x-iy$. Using this notation, the left and right representations can be written
$$l(a)=\begin{pmatrix} l_M(a)&0\\0&l_M(\overline a)\end{pmatrix},\quad r(a)=\begin{pmatrix} r_M(\overline a)&0\\0&r_M(a)\end{pmatrix}.$$

Since $\mathcal A\cong M_n(\C)$, the unitary elements of the algebra form the group $\U(n)$. The gauge group action is the adjoint action
\begin{equation}l(a)r(a^{-1})=\begin{pmatrix} l_M(a)r_M(\overline a^*)&0\\0&l_M(\overline a)r_M(a^*)\end{pmatrix},
\label{cxgaugegroup}\end{equation}
from which it follows that the gauge group of the spectral triple is
\begin{equation*}
    G(\mathcal{A},\mathcal{H};J) = \U(n)/\mathbb{Z}_2,
\end{equation*}
in agreement with \cite{Bhowmick2011QuantumGeometry}. 

The subgroup of unitary elements $u\otimes1$, for $u\in\mathcal A_M$, determine a $\OO(n)/\Z_2$ or $\Sp(n/2)/\Z_2$ subgroup of $G(\mathcal{A},\mathcal{H};J)$ that is the same as the gauge group of the matrix geometry without the internal space. Moreover, these transformations act identically on the two components of $\mathcal H_F$ and so are universal (the same for all particles). Thus these are the geometric gauge transformations for this model. Another subgroup is the set of elements $1\otimes \kappa\in\mathcal A$, which determine the $\U(1)/\Z_2\cong\U(1)$ gauge group of the internal space. These gauge transformations act as 
$$\begin{pmatrix} \kappa^2&0\\0&\kappa^{-2}\end{pmatrix}$$
and so depend on the internal space charges. 

The analogue of the space-time dependent gauge transformations can be understood by considering $\mathfrak U(\mathcal A)$, the Lie algebra of $\U(\mathcal A)$. An element $a=x+iy\in \mathfrak U(\mathcal A)$ has $x^*=-x$, $y^*=y$. The splitting into real and imaginary parts is a $\Z_2$ grading of $\mathfrak U(\mathcal A)$, and in particular, the commutator of two imaginary elements is real.

The adjoint action of $a$ is
\begin{equation}l(a)+r(a^*)=\begin{pmatrix} [x,\cdot\,]+i\{y,\cdot\,\}&0\\0&[x,\cdot\,]-i\{y,\cdot\,\}\end{pmatrix}.
\label{gaugeliealg}\end{equation}
The y terms are the non-commutative analogue of the standard notion of a gauge transformation on a commutative manifold: the anti-commutator with $y$ is the analogue of multiplication of a fermion by the gauge potential, with sign depending on the corresponding gauge charge. This anticipates the generation of gauge fields by fluctuations in \S\ref{sec:fluct}. 

Note that two successive imaginary transformations generate a real transformation, corresponding to the transformation of the manifold in \S \ref{sec:fluctuations}. This is analogous to successive gauge transformations generating a spacetime transformation. This behaviour is related to that observed in non-commutative field theory, as described in \cite{Szabo2001QuantumSpaces}. There, the presence of the star product deforms the gauge symmetry so that gauge transformations become intrinsically non-local, involving an infinite series of derivative corrections. As a consequence, the distinction between internal gauge transformations and spacetime symmetries is blurred.

\section{Fluctuations of an (0,4) Matrix Geometry with U(1) Internal Space}\label{sec:fluct}

The vacuum Dirac operator takes the form
\begin{equation}
    D_0 = (\diracgamma  \comm{L_j}{\cdot} + \trigamma{j}{k}{l}  \acomm{H_{jkl}}{\cdot}) \otimes 1_F
\label{vacDirac}\end{equation}
where the sum over repeated indices is taken in the same manner as \eqref{eq:Dirac}. Connes one-forms for this model are given by 
\begin{equation*}
        \omega = \sum_m l(a_m)\comm{D_0}{l(b_m)}
\end{equation*}
where $a_m,b_m\in\A$. 

Using the fluctuations of matrix geometries given by \eqref{leftfluc}, and denoting $\Lambda_j = \sum_m a_m\comm{L_j}{b_m}$, $\Lambda_{jkl} = \sum_m a_m\comm{H_{jkl}}{b_m}$, the left fluctuations are  
\begin{equation*}
    \omega = \diracgamma \otimes \begin{pmatrix}
            l_M(\Lambda_j)  & 0 \\ 0 & l_M(\overline{\Lambda_j}) 
        \end{pmatrix} 
        + \trigamma{j}{k}{l} \otimes \begin{pmatrix}
            l_M(\Lambda_{jkl})  & 0 \\ 0 & l_M(\overline{\Lambda_{jkl}}) 
        \end{pmatrix}
\end{equation*}
As even geometries have $\varepsilon'=1$, gamma matrices and the real structure commute.
The total fluctuations are thus
\begin{align*}
    \Omega &= \diracgamma \otimes \begin{pmatrix}
        l_M(\Lambda_j) + r_M(\overline{\Lambda_j^*}) & 0 \\ 0 & l_M(\overline{\Lambda_j}) + r_M(\Lambda_j^*) 
    \end{pmatrix} \\
    &+ \trigamma{j}{k}{l} \otimes \begin{pmatrix}
        l_M(\Lambda_{jkl}) + r_M(\overline{\Lambda_{jkl}^*})  & 0 \\ 0 & l_M(\overline{\Lambda_{jkl}}) + r_M(\Lambda_{jkl}^*)
    \end{pmatrix}
\end{align*}

\subsection{Real and imaginary fluctuations}
There are two types of fluctuations related to where in the algebra, $\A = M_n(\C)$, that $\Lambda_j$ and $\Lambda_{jkl}$ reside. To see this, decompose the matrices into $\Lambda_j= \sigma_j + i \theta_j$ and $\Lambda_{jkl}= x_{jkl} + i y_{jkl}$ where $\sigma_j,y_{jkl} \in \A_M$ are anti-Hermitian and $\theta_j,x_{jkl} \in \A_M$ are Hermitian.

The first type are the \emph{real fluctuations} where the matrices are contained in the real subgroup, i.e., $\theta_j = y_{jkl} = 0$.
For these fluctuations, the $2\times 2$ matrix factor is the identity matrix and the one-forms are generated solely by the `manifold' algebra $\mathcal A_M$. The total fluctuation takes the same form as for the original Dirac operator $D_M$, and is given by
\begin{align*}
            \Omega_\R &= \left(\diracgamma \comm{\sigma_j}{\cdot} + \trigamma{j}{k}{l}  \acomm{x_{jkl}}{\cdot} \right) \otimes 1_F\\
            &= (\Sigma + X) \otimes 1_F.
\end{align*}
For later convenience, the notation $\Sigma = \diracgamma  \comm{\sigma_j}{\cdot}$ and $X = \trigamma{j}{k}{l} \acomm{x_{jkl}}{\cdot}$ is used. 

The second type are the \emph{imaginary fluctuations} where $\sigma_j = x_{jkl} = 0$. For these fluctuations, the structure of the internal space becomes important. The total fluctuation is given by
\begin{align*}
           \Omega_{i\R} &= \left(\diracgamma  \acomm{i\theta_j}{\cdot}  + \trigamma{j}{k}{l}  \comm{iy_{jkl}}{\cdot} \right) \otimes \Gamma_F \\
           &= (\Theta + Y) \otimes \Gamma_F,
\end{align*}
using the notation $\Theta = \diracgamma \acomm{i\theta_j}{\cdot} $ and $Y = \trigamma{j}{k}{l}  \comm{iy_{jkl}}{\cdot} $.

\subsection{Physical Analogies for Fluctuations}
To provide a physical interpretation of these non-commutative fluctuations, consider first the almost commutative case. Here, the total automorphism group splits as a semi-direct product of the manifold diffeomorphisms (the outer automorphisms) and the internal symmetry (the inner automorphisms). Connes' one-forms, and in turn the inner fluctuations, generate the gauge fields associated to internal symmetries, whilst the outer automorphisms are not associated with fluctuations.

For a spectral triple with a matrix geometry as the `manifold' this is no longer true. The manifold algebra is a simple matrix algebra, thus by the Skolem-Noether theorem, all the automorphisms are inner. As a result, Connes' one-forms can generate more than just gauge fields. In the case with the internal space, both real and imaginary fluctuations are generated. These fluctuations can be understood by extending the analogy with commutative geometry for the gauge transformations discussed in \S \ref{sec:internal}.

The real fluctuations are identical to those of an (0,4) matrix geometry without an internal space. As a result of this, the discussion in \S \ref{sec:fluctuations} applies in this case. Redefining the manifold Dirac operator,
\begin{equation*}
    D_M' \equiv D_M + \Sigma + X = \diracgamma  \comm{L_j + \sigma_j}{\cdot} + \trigamma{j}{k}{l}  \acomm{H_{jkl} + x_{jkl}}{\cdot},
\end{equation*} 
it is clear that the real fluctuations are perturbations of the underlying manifold.

The imaginary fluctuations contain an anti-commutator term generated by $L_j$, and a commutator term generated by $H_{jkl}$. In addition, these terms are charged under $\U(1)$ on the internal space, as ($\pm i$). These fluctuations cannot be absorbed via a redefinition of the manifold Dirac operator and thus need an alternative interpretation. 

The first term, $\diracgamma\acomm{i\theta_j}{\cdot}\otimes \Gamma_F$, admits a straightforward interpretation. It is multiplicative and charged under the gauge group. This is analogous to a Lie algebra-valued function of spacetime, or, in other words, a gauge field.
The second term, $\trigamma{j}{k}{l} \comm{iy_{jkl}}{\cdot} \otimes \Gamma_F$ is more complicated. As stated, a commutator is the non-commutative analogue for a derivative operator. Hence, this is analogous to a charged derivative operator.

A further insight into the geometric character of the fluctuations can be gained by calculating the gauge transformations \eqref{cxgaugegroup}. The subgroup of real group elements, $\U(\mathcal A_M)$, preserves the type of each term in the Dirac operator,
$$ [L,\cdot\,]\otimes K \mapsto [uLu^*,\cdot\,]\otimes K$$
$$ \{H,\cdot\,\}\otimes K \mapsto \{uHu^*,\cdot\,\}\otimes K$$
for either $K=1_F$ or $K=\Gamma_F$.

The imaginary elements of $\U(\mathcal A)$ don't form a subgroup, so it is more informative to explore the transformation associated to imaginary elements of the Lie algebra, $iy\in\mathfrak U(\mathcal A)$, as in \eqref{gaugeliealg}. 
These give the infinitesimal gauge transformations
$$ [L,\cdot\,]\otimes K \mapsto \{[iy,L],\cdot\,\}\otimes K\Gamma_F$$
$$ \{H,\cdot\,\}\otimes K \mapsto [[iy,H],\cdot\,]\otimes K\Gamma_F$$
In geometric terms, a universal derivative becomes a charge-dependent function (as it does in the usual gauge theory), but also a universal function becomes a charge-dependent derivative. Under successive transformations, the reverse process can occur with charge dependent terms becoming universal.

The gauge transformations apply equally well to the terms in the vacuum Dirac operator \eqref{vacDirac} as to the fluctuations. The gauge transformations mix the fluctuations $\Sigma\otimes 1$ with $\Theta\otimes \Gamma$, and separately mix the fluctuations $X\otimes 1$ with $Y\otimes \Gamma$. Since the $\Sigma$ and $\Theta$ terms together are understood to be the non-commutative analogues of fluctuations of a gauge-covariant Dirac operator, it begs the question of the interpretation of the other pair. A possible answer is that these two pairs are related by the \emph{chiral rotation} \cite{Barrett2015MatrixTriples} determined by the operator $R=e^{i\pi\Gamma/4}$, which transforms a Dirac operator $D$ into a new Dirac operator
$$\widetilde D=RDR^{-1}=iD\Gamma.$$
Applying this to the fluctuations gives 
$$R(X\otimes 1_F)R^{-1}=\eta\, \gamma^m \{ix_{jkl},\cdot\,\}\otimes \Gamma_F,$$
with $m,j,k,l$ all different and $\eta=\epsilon^{lkjm}=\pm1$.
This is of the same type as $\Theta\otimes\Gamma_F$. Similarly,
$$R(Y\otimes \Gamma_F)R^{-1}=\eta\, \gamma^m [-y_{jkl},\cdot\,]\otimes 1_F,$$ which is of the same type as $\Sigma\otimes 1_F$.

\section{The Real Fermion Integral}\label{sec:ferm}
\subsection{The Total Dirac Operator and the Fermion Action}
The Euclidean fermion action is
\begin{equation*}
    S = \frac12\inprod{J\Psi}{D\Psi}
\end{equation*}
where $\Psi\in\hilbert$, and $D= D_0 + \Omega_{i\R} + \Omega_\R$ is the fluctuated Dirac operator. The vacuum Dirac operator, $D_0$, describes the dynamics of a free fermion and, in conjunction with the fluctuations, determines the interactions of fermions with the bosonic fields in a given theory \cite{vanSuijlekom2025NoncommutativePhysics}.

The total fluctuated Dirac operator for this model is
\begin{equation}
\begin{aligned}
        D &= \begin{pmatrix}
            (D_M + \Sigma + X) + (\Theta + Y) & 0 \\
            0 & (D_M + \Sigma + X) - (\Theta + Y)
        \end{pmatrix} \\
        &= \begin{pmatrix}
            D_M' + (\Theta + Y) & 0\\
            0 & D_M'- (\Theta + Y)
        \end{pmatrix}
\end{aligned}\label{flucD}
\end{equation}
The action for a fermion field, $\Psi=\begin{pmatrix} \chi \\ \xi \end{pmatrix} \in \hilbert$, is
\begin{align*}
   \frac12 \inprod{J\Psi}{D\Psi} &=\frac12 \inprod{J_M\xi}{(D_M' + (\Theta + Y))\chi}  \\&+\frac12\inprod{J_M\chi}{(D_M' - (\Theta + Y))\xi}
\end{align*}

The operator $D_M'$ is universal (chirality-neutral). This operator acts as the analogue of the standard Dirac operator in a commutative theory. In comparison to $D_M$, this operator has a modified spectrum. Consequently, the kinetic term of the action experiences a finite shift due to the fluctuation of the geometry, and the fermion propagator undergoes a corresponding shift.  
The charged term originating from the anti-commutator, $\pm \Theta$, provides a gauge field-like interaction. In addition, the charged term generated by the commutator, $\pm Y$, provides a novel derivative-like term acting on the spinor fields. This arises purely due to the non-commutative structure of the fluctuations.

\subsection{Fermionic Integrals for Spectral Triples}
The fermionic action can be used, together with a bosonic spectral action, to form a functional integral over both the field content and the space of geometries \cite{Hessam2022FromTheory}. This approach provides a method to compute expectation values for observables and thus quantise the theory defined by a spectral triple. 

Due to the finite dimensional nature of matrix geometries, fermions can be integrated out exactly for certain models, resulting in a closed-form expression. Integrating out these fields produces an effective partition function for the bosonic fields. The fermion integral is defined by 
\begin{equation*}
    Z  = \int_{\mathcal{H}} \mathcal{D}\Psi\, e^{\frac{i}{2}\langle J\Psi, D \Psi \rangle }
\end{equation*}
The real structure in the inner product ensures this integration takes place only over the fermion field, $\Psi$. There is no need to introduce an independent conjugate field. 

The standard expression for this integral is a Pfaffian. In the case of KO-dimension 2, the Pfaffian can be evaluated using one of the canonical bases determined by $J$ \cite{Barrett2024FermionTriples}. The result is non-negative and is given by  
\begin{equation*}
    Z = \sqrt{\det(D)}.
\end{equation*}

Integrating out fermions generates fermionic loop corrections to the bosonic action, analogous to standard QED \cite{Peskin2018AnTheory}. Real fluctuations of the vacuum Dirac operator, $D_M \otimes 1_F$, shift the vacuum to a new operator, $D'_M \otimes 1_F$, for which the integral is 
\begin{equation*}
    Z = \sqrt{\det(D'_M)^2}=\det(D'_M).
\end{equation*}
This result can also be obtained by regarding $\xi$ as the field conjugate to $\chi$, and calculating the standard (complex) fermion integral for $D'_M$. The more interesting case of the imaginary fluctuations is considered in the next section.

\subsection{Fermion integral for imaginary fluctuations}
The determinant for the fluctuated Dirac operator \eqref{flucD} can be expressed as  
     \begin{equation*}
        \det(D) =  \det(D_M'^2 -\underbrace{( ( \Theta + Y)^2 + \comm{D_M'}{ \Theta + Y})}_{F})
     \end{equation*}
The terms induced by the fluctuations take the form
\begin{equation*}
    F=  \comm{D_M'}{ \Theta + Y}+(\Theta + Y)^2 
\end{equation*}
Under the standard spectral triple identification of $\comm{D}{\cdot}$ as the differential $d$, the above operator resembles the usual field strength 2-form for a commutative theory
\begin{equation*}
    F = dA + A^2
\end{equation*}
A similar field strength term for the case of a complex matrix geometry is derived via the spectral action in \cite{Perez-Sanchez2021OnModel}.

Upon expanding the total field strength 
     \begin{align*}
         F &=   \comm{D_M'}{\Theta} +\Theta^2+   \comm{D_M'}{Y} +Y^2+ \acomm{\Theta}{Y} \\
         &= F_\Theta + F_Y +  \acomm{\Theta}{Y},
     \end{align*}
it can be seen that there are individual field strength-like terms for both $\Theta$ and $Y$, and also a mixing term. The term $F_\Theta$ is the standard field strength expression for the gauge field $\Theta$, but $F_Y$ is something new, since $Y$ is itself a derivative operator.

The resulting fermionic determinant defines an effective action via a perturbative expansion. Expanding in powers of $F$ produces fermion loop diagrams with external field-strength insertions. In addition, there is a new interaction vertex coupling a fermion simultaneously to both fluctuation fields, originating from the mixing term. 

\printbibliography

\end{document}